\documentstyle[aps,multicol,epsfig]{revtex}
\begin{document}
\draft
\title{ Stochastic Resonance in Ion Channels Characterized by Information
Theory}
\author{Igor Goychuk and Peter H\"anggi}
\address{Institute of Physics, University of Augsburg,
Universit\"atsstrasse 1,
86135 Augsburg, Germany}
\author{}
\address{}
\maketitle
\widetext
\vspace{-1cm}
\begin{abstract}

We identify  a unifying   measure for stochastic resonance (SR) in voltage
dependent ion channels which comprises  periodic (conventional), aperiodic 
and  nonstationary SR. Within a simplest setting, the gating dynamics is 
governed by two-state conductance fluctuations,  which switch at random time
points between two values. The corresponding continuous time point process is
analyzed by virtue of information theory. In pursuing this goal we evaluate for
our dynamics  the $\tau$ -information, the mutual information and the rate of
information gain. As a main result we  find  an analytical formula for the rate
of information gain that solely involves the probability  of the two channel
states and their noise averaged  rates. For small voltage  signals it
simplifies to a    handy expression. Our findings are applied to study SR in a
potassium channel. We find that SR occurs only  when the closed state is
predominantly dwelled. Upon increasing the probability for the open channel
state the application of an extra dose of noise  monotonically deteriorates the
rate of information gain, i.e., no SR behavior occurs.\\\

\end{abstract}

\raggedcolumns
\begin{multicols}{2}
\narrowtext
\noindent

\section{Introduction}

The stochastic resonance (SR) constitutes a cooperative phenomenon wherein the
addition of noise to the information carrying signal can improve in a
paradoxical manner the detection and transduction of signals in nonlinear
systems (see, e.g., \cite{wies} for an introductory overview and \cite{rev1}
for a comprehensive survey and references). Clearly, this effect could play a
prominent role for the function  of sensory biology. As such, the beneficial 
role of ambient and external noises has been addressed not only theoretically
(see, e.g., \cite{longtin}),  but also has been manifested experimentally on
different levels of biological organization -- e.g., in the human visual
perception \cite{sim} and tactile sensation \cite{col1}, in the cricket cercal
sensory systems  \cite{levin}, or also in the mammalian neuronal networks
\cite{gluck} and -- even  earlier -- for the mechanoreceptive system in
crayfish \cite{doug}. Presumably, the molecular mechanisms of the biological SR
have their roots in stochastic properties of the ion channel arrays of the
receptor cell membranes \cite{wies}. This stimulates the interest to study SR
in biological ion channels. One of the outstanding challenges in SR-research
therefore is the quest to answer whether -- and how -- SR occurs in single
and/or coupled ion channels.

These channels are the  evolution's solution of enabling membranes made out of
fat to participate in electrical signaling. They are formed of special membrane
proteins \cite{hille}. In spite of the great diversity, these natural occurring
nanotubes share some common features. Most importantly, the channels are
functionally bistable, i.e. they are either {\it open}, allowing specific ions
to cross the membrane,  or are {\it closed}\cite{hille}. The regulation of the
ion flow is achieved by means of the so-called gating dynamics, i.e., those
intrinsic stochastic transitions occurring inside the ion channel that regulate
the dynamics of open and closed states. The key feature of gating dynamics is
that the opening-closing transition rates depend strongly on  external factors
such as the membrane potential (voltage-gated ion channels), membrane tension
(mechanosensitive ion channels), or presence of chemical ligands (ligand-gated
ion channels). This sensitivity allows one to look upon the corresponding ion
channels as a kind of single-molecular sensors which transmits an input
information   to the signal-modulated ion current response.

Recently, it has been demonstrated experimentally  by Bezrukov and Vodyanoy
\cite{bez} that a parallel ensemble of independent, although {\it artificial}
(alamethicin) voltage-gated ion channels does exhibit SR behavior,  when the
information-carrying voltage signal is perturbed by a noisy component.  These
authors have put forward the so-called {\it non-dynamical model} of SR. It is
based on the statistical analysis of the "doubly stochastic", periodically
driven Poisson process with corresponding voltage-dependent spiking rate
\cite{bez,bez2}. Conceptually, such a model can be adequate to  those
situations only where the channel is closed on average with openings
constituting relatively rare events. An experimental  challenge is to verify
whether the  SR effect persists for {\it single} natural biological ion
channels under realistic conditions. Moreover, a second challenge is to extend
the theoretical description in \cite{bez2} to account properly for the
distribution of dwell times  spent by the channel in the conducting state.

The previous research on SR in ion channels has exclusively been restricted  to
the case of conventional SR, i.e., SR with a periodic input signal. In a more
general situation, however,  input aperiodic signals  can be drawn from some
statistical distribution.   This case of the so-termed  {\it aperiodic} SR has
recently been put forward for neuronal systems
\cite{levin,collins,stemmler,buls}. Note that the important assumption of
dealing with a  signal realization that is taken from a stationary process has
been made in all previous studies. In practice, however, one frequently meets 
a situation where this stationarity assumption is not rigorously valid, because
the signal has a finite duration on the time-scale set by observation. In this
{\it nonstationary} situation, both the spectral and the  cross-correlation SR
measures are inadequate. A preferable approach is then to look for SR from the
perspective of statistical information transduction \cite{levin,buls}.  As is
elucidated with this work, information theory \cite{shannon} can indeed provide
a {\it unified} framework to address different types of  SR, including  {\it
nonstationary} SR. It is the main purpose of this work to investigate  the
possibility to enhance the transmission of information in a {\it single} ion
channel in presence of a dose of noise. This task will be accomplished  within
a simplistic two-state Markovian model for the ion channel conductance
\cite{hille}. Already within such an idealization, our analysis in terms
of information theory measures turns out to be rather involved.

\section{Two-state model}

In principle, the microscopic description of the gating dynamics should be
based upon the detailed understanding of the structure of channel's ``gating
dynamics''. Present state of the art assumes  that the voltage-sensitive gates
are represented by mobile charged  $\alpha$~-~helix fragments of the channel
protein which can dynamically  block the ion conducting pathway. Therefore, the
gating dynamics can be described by diffusive motion of gating ``particles'' in
an effective potential. Then, Kramers diffusion theory \cite{kramers,hanggi1}
and its extension to the realm of {\it fluctuating barriers} (see, e.g.,
\cite{hanggi2} for a review and further references) can be utilized to describe
the gating dynamics. Such a type of procedure, however, is still in its  
infancy \cite{bezanilla}. For our purpose,  it  suffices to follow a
well-established  phenomenological road provided by a discrete phenomenological
modeling \cite{sak}.

\begin{figure}[h]
\begin{center}
\leavevmode
\epsfxsize=0.45\textwidth
\epsfbox{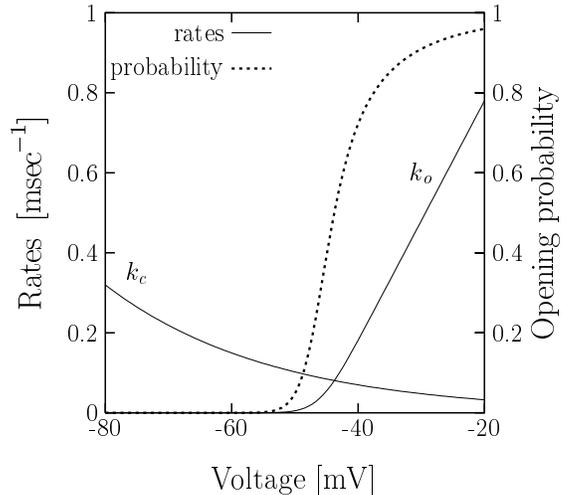}
\end{center}
\caption{Voltage dependence of the opening rate, $k_o$, and the closing
rate, $k_c$,
for a K$^+$ ion channel {\it vs.} a
static voltage $V_0$ (solid lines), cf. Eq. (A1).
The corresponding probability for the channel in the open state is depicted
by the dotted line. }
\label{Fig1}
\end{figure}

The simplest two-state model of this kind reflects the functional bistability
of ion channels. The dichotomous fluctuations between the conducting and
nonconducting conformations of {\it single} ion channels are clearly seen in
the patch clamp experiments \cite{sak}. The statistical distributions of
sojourn times of the open channel state and the closed channel state,
respectively, are generically not exponentially distributed \cite{sak}.
However, one can  characterize these  time distributions by its average,
$<T_o(V)>$, to dwell  the open (O) state, and by  its corresponding average,
$<T_c(V)>$, to stay in the  closed (C) state. These two averages  depend on the
transmembrane voltage $V$.  Then, the actual multistate gating dynamics can be
approximately mapped onto the effective two-state dynamics described by the
simple kinetic scheme \begin{eqnarray*}\\ C\begin{array}{c}  
k_o(V) \\[-1.9em] \longrightarrow
\\[-1.5em] \longleftarrow \\[-1.9em] k_c(V) \end{array} O \\
\end{eqnarray*} with corresponding
voltage-dependent effective transition rates $k_{c}(V)=1/<T_o(V)>$ and
$k_{o}(V)=1/<T_c(V)>$, respectively. Although such a two-state Markov
description presents a rather crude approximation, it captures the main
features of gating dynamics of the voltage-sensitive ion channels  -- the
dichotomous nature and the voltage-dependence of transition rates. Moreover, 
this model yields by construction the correct mean open (closed) dwell times,
and the stationary probability for the channel to stay open, i.e. 
$P_o(V)=<T_o(V)>/(<T_o(V)>+<T_c(V)>)$. An example for the experimental
dependence of the transition rates on voltage $V$ can be found for a $K^{+}$
channel in Ref. \cite{salman,marom} and is depicted in Fig. \ref{Fig1}. We note
that in contrast to the closing rate the {\it effective} opening
rate has {\it no  exponential} dependence on the
voltage. Thus, these two rates are not symmetric 
(with respect to dependence on $V$, cf. Fig. 1).
The reason being that the two-state description results as the {\it
reduction} of an intrinsic multistate (or multi-well) gating dynamics and thus
presents only a shadow of real behavior. In this sense, the Markovian
approximation models the true non-Markovian dynamics on a coarse grained time
scale.

To proceed, one has to generalize this working  model to the case with
time-dependent voltages $V(t)=V_0+V_s(t)+V_n(t)$. Here we distinguish among
three components of the voltage: (i) the constant bias voltage $V_0$, (ii) some
time-dependent, unbiased signal $V_s(t)$, and (iii) a noisy component voltage
$V_n(t)$. The noisy voltage $V_n(t)$ is assumed to be a stationary Gaussian
Markovian noise with zero average and root mean squared (r.m.s.) amplitude
$\sigma$. Moreover, it possesses a frequency bandwidth $f_n$. Let us restrict
our treatment to the situation where {\it both} the signal and the external
noise are slowly varying on the time-scale set by diffusive motions occurring
within the open (or closed) conformation. This time-scale $\tau_{con}$
typically lies  in the $\mu{\rm sec}$ range as manifested experimentally by the
fast events in  channel activation \cite{bezanilla}. We thus can apply the {\it
fluctuating rate} model \cite{wies,bez2} assuming that the transition rates
$k_{o(c)}(t)\equiv k_{o(c)}[V(t)]$ follows {\it adiabatically}  the voltage
$V(t)$. Furthermore, we assume that the applied Gaussian voltage $V_n(t)$
presents effectively  ``white-noise" on the time scale set by the decay of
autocorrelations  of the ion current fluctuations. The autocorrelation time
$\tau_I=1/[k_o(V_0)+k_c(V_0)]$ is  typically of the order of milliseconds
\cite{sak}. Then, the  choice of a noise bandwidth $f_n$ satisfying
$\tau_I^{-1}\ll f_n\ll\tau_{con}^{-1}$, i.e., $f_n \sim 10-100 \;{\rm kHz}$,
presents a  consistent specification for the fluctuating rate description. The
role of external noise is thus reduced within the same two-state approximation
merely to forming new, noise-dressed time-dependent transition rates $\bar
k_{\alpha=o,c}(t)\equiv\langle k_{\alpha=o,c}[V(t)]\rangle_n$. They result by
taking the stochastic average of the fluctuating rates over the {\it external}
noise. These effective rates depend now on the noise r.m.s. amplitude $\sigma$,
the static voltage $V_0$ and the time-dependent signal $V_s(t)$. It turns out
that within in the given approximation the averaged transition rates do not
depend on the noise bandwidth $f_n$, see also Appendix A.

Our model for the channel dynamics thus reads

\begin{eqnarray} \label{balance}
\frac{d P_{o}(t)}{dt}=-\bar k_{c}(t)P_{o}(t)+\bar k_{o}(t)P_{c}(t), \nonumber\\
\frac{d P_{c}(t)}{dt}=-\bar k_{o}(t)P_{c}(t)+\bar k_{c}(t)P_{o}(t),
\end{eqnarray}
where $P_o(t)$, and $P_c(t)$ denote the time-dependent probabilities
for a single  ion channel to be open or closed, respectively. The
stochastic process
described by Eq. (\ref{balance}) is a {\it nonstationary} random telegraph
noise (RTN) with  time-dependent transition rates. This model
has extensively been studied in the literature, for example, to
model conventional SR \cite{rev1,mc}.  Moreover, this model has been
studied in Ref. \cite{neiman}
from the perspective of
input-output cross--correlations as a simple model for {\it aperiodic} SR.
However, to the best of our knowledge the detailed 
analysis of this cornerstone
model from the perspective to use   information theory \cite{shannon,buls} to
specify the information transduction process
 is still lacking.

\section{Statistical distribution of current fluctuations}

How can we estimate the amount of information transmitted from the input
voltage signal $V_s(t)$ to the output ion current $I(t)$? A comparative
statistical analysis of the ion current fluctuations performed in the absence
and in the presence of signal allows one to answer this question.

When the channel is open a large number of ions cross the channel; thus
creating a finite current $I_o(t)$. This current obeys the Ohmic law,
$I_o(t)=g_o [V(t)-V_k]$, where $g_o$ is the conductivity of the open channel
and $V_k$ is the ``reversal'' potential (Nernst potential) for $K^{+}$ ion
flow. When the channel is closed, the ion flow is negligible and the current is
zero. We recall that the current passing through the  open channel is generally
time-dependent in accordance with the externally applied signal $V_s(t)$.
However, we will assume that the information about signal is encoded in the
switching events of current between zero and $I_o(t)$, and {\it not} in the
additional modulation of $I_o(t)$. In other words, the information is assumed
to be encoded in the signal-modulated {\it conductance} fluctuations between
$g_o$ and zero \cite{remark1}.

Moreover, one can describe the resulting current fluctuations in terms of
conductance fluctuations, i.e., \begin{equation} \label{current} I(t) = g(t)
[V(t) - V_k] \end{equation} wherein $g(t)$ as a two-state random point process
\cite{kamp,strat}. The sample space of $g(t)$ within the time interval $[0,t]$
consists of stochastic trajectories which flip between zero and $g_0$ at
randomly distributed switch-time points $\tau_i$, $i$ = 1,2,...; i.e.,
\begin{eqnarray}\label{tdomain}
0<\tau_1<\tau_2<...<\tau_s<t .
\end{eqnarray}
This defines a continuous time point process $\tau(\tilde t)$, $0\leq \tilde t 
\leq t$. Next, we divide the sample space into  two subspaces: (i) the subspace
``o'' contains all trajectories which finish in the open state at the end point
$t$ of the considered time interval, and (ii) the subspace ``c'' which contains
all trajectories which end in the closed state, respectively. Furthermore,
within each subspace the trajectories are divided into the subclasses described
by the number $s=0,1,2,...$ which enumerates the number of intermediate flips
that occurred between  open and closed states in order to arrive at the final
state. The probability distribution on this  space is given by a sequence of
joint multi-time probability densities $Q^{c(o)}_s(t,\tau_s,..,\tau_1)$  for
switches to occur at time $\tau_1, \tau_2, ....,\tau_s$ and  to end up  at time
$t$ in either the open state $o$ or closed state $c$, respectively. This
probability distribution is normalized; i.e.,
\begin{eqnarray}\label{norm}
\sum_{\alpha=o,c}[Q_0^{\alpha}(t)+\sum_{s=1}^{\infty}
\int_{0}^{t}d\tau_s && \int_{0}^{\tau_s}d\tau_{s-1}...\nonumber \\
\times && \int ^{\tau_2}_{0}d\tau_1
Q^{\alpha}_s(t,\tau_s,..,\tau_1)]=1\;.
\end{eqnarray}
The probability densities $Q^{c(o)}_s(t,\tau_s,..,\tau_1)$ are readily
constructed by taking into account the facts that the process $g(t)$
is (semi-)Markovian
for any given realization of the voltage signal  $V_s(t)$
with the switching time points $\tau_i$ being drawn alternatingly from
two different {\it time-dependent} Poisson distributions \cite{strat}.
In particular, the probability to stay in the closed conformation until
time $t$, given that this conformation has been occupied initially with the
probability $P_c(0)$, is
\begin{eqnarray}\label{Q0}
Q_0^{c}(t)=e^{-\int_0^{t}\bar k_{o}(\tau)d\tau}P_c(0)\;.
\end{eqnarray}
To obtain the remaining  probability densities, we introduce
the conditional probability density
\begin{eqnarray}\label{cond}
P_{c}(\tau_2|\tau_1)=\bar k_{o}(\tau_2)e^{-\int_{\tau_1}^{\tau_2}
\bar k_{o}(\tau)d\tau}\;
\end{eqnarray}
for leaving the state $``c"$  in the
time interval  $[\tau_2+dt,\tau_2]$ given that this state
was occupied with  probability one at $t=\tau_1$. Analogous expressions, with
indices changed from $``c"$ to $``o"$, hold
obviously also for the complementary quantities
$Q_0^{o}(t)$ and $P_{o}(\tau_2|\tau_1)$. Then,
the multi-time probability densities emerge as
\begin{eqnarray}\label{qeven}
Q_{2n}^{c}(t,\tau_{2n},...,\tau_1)=
e^{-\int_{\tau_{2n}}^{t}\bar k_{o}(\tau)d\tau}
P_{o}(\tau_{2n}|\tau_{2n-1})\nonumber \\ \times 
P_{c}(\tau_{2n-1}|\tau_{2n-2})...
P_{o}(\tau_{2}|\tau_{1})P_{c}(\tau_{1}|0)P_c(0) \;,
\end{eqnarray}
for a given even number of flips, and
\begin{eqnarray}\label{qodd}
Q_{2n+1}^{c}(t,\tau_{2n+1},...,\tau_1)=e^{-\int_{\tau_{2n+1}}^{t}\bar k_{o}
(\tau)d\tau}P_{o}(\tau_{2n+1}|\tau_{2n}) \nonumber \\ \times 
P_{c}(\tau_{2n}|\tau_{2n-1})...
P_{c}(\tau_{2}|\tau_1)P_{o}(\tau_{1}|0)P_o(0) \;,
\end{eqnarray}
for odd number of flips, respectively.
The probability densities for  the other subspace ending in the open state
(labeled with $``o"$)
can be written down by use of a  simple interchange of the indices $``c"$ and
$``o"$
in Eqs. (\ref{Q0}) - (\ref{qodd}).

The above  reasoning yields a {\it complete} probabilistic description of
the  stochastic switching
process that is related to the conductance  fluctuations $g(t)$.
 In terms of the stochastic path description,
the probability that the channel is open
at the instant time $t$ is therefore given by
\begin{eqnarray}\label{solution}
P_{o}(t)=  Q_0^{o}(t) + \sum_{s=1}^{\infty}
\int_{0}^{t}d\tau_s && \int_{0}^{\tau_s}d\tau_{s-1}...\nonumber \\ \times
&& \int^{\tau_2}_{0}d\tau_1
Q^{o}_s(t,\tau_s,..,\tau_1) \;.
\end{eqnarray}
An analogous expression holds also for the probability of the closed
conformation
$P_c(t)$.  Upon differentiating $P_o(t)$ and $P_c(t)$ with respect to time $t$
one can check that these time-dependent probabilities indeed satisfy the
kinetic equations (\ref{balance}).

\section{ Stochastic resonance quantified by information theory }

In the following we derive the general theory for various information
measures that can be used to quantify the information gain obtained from an
input signal $V_s(t)$ being transduced by  the ion channel
current realizations $I_i(t)$ when $V_s(t)$ is switched on, versus the case
with  $V_s(t)$
being switched off.  Intuitively, this information describes the difference
in uncertainty
about the current realizations in the absence and in the presence of the
signal $V_s(t)$.
\subsection{Preliminaries}
We start out by reviewing the necessary background. Let us first consider a
{\it discrete random
variable} ${\cal A}$.
As demonstrated by K. Shannon in 1948 \cite{shannon}
(his expression was
discovered independently by N. Wiener), the information entropy

\begin{eqnarray}\label{sh1}
 S({\cal A})=- \kappa\sum_{i=1}^n p_i\ln p_i\;
\end{eqnarray}
provides a  measure for the uncertainty about a particular realization $A_i$
of ${\cal A}$ \cite{remark}. In Eq. (\ref{sh1}), the set ${p_i}$ denotes the
normalized
probabilities for the realizations $A_i$ to occur, $\sum_{i=1}^n p_i=1$.
The positive
constant $\kappa$ in (\ref{sh1}) defines the unit used in measurement. If the
information entropy is measured  in binary units, then $\kappa=1/\ln 2$,
natural
units  yield $\kappa=1$, and digits give $\kappa=1/\ln 10$. This measure
attains a minimum (being zero) if and only if one $p_i=1$ for a particular
value of $i$, and all others satisfying $p_i=0$. It reaches a maximum if $p_i=1/n$.
The information entropy for a probability distribution is therefore a measure
of how strongly it is peaked about a given alternative.
The {\it uncertainty}  is consequently large for spread out distributions
and small for concentrated ones.

The application of an external
signal (perturbation) results in a change of probabilities $p_i$ and
consequently in entropy
$S({\cal A})$. The gained information  ${\cal I}$  is then defined by the
corresponding change in entropy, i.e.  ${\cal I}=S_{before}-S_{after}$.

The generalization of the information concept onto the case of continuous
variable
$A(x)$ presents no principal difficulties. In this
case a proper definition of entropy reads
\begin{eqnarray}\label{sh2}
 S(A) & = &- \kappa\int p(x)\ln [p(x)\Delta x]dx \nonumber \\
 & \equiv & -\kappa\int p(x)\ln [p(x)]dx-\kappa\ln \Delta x;
\end{eqnarray}
wherein $p(x)$ is the probability density and $\Delta x$ denotes the precision
with which the variable $A(x)$ can be measured (coarse graining of cell
size). As is clearly seen from Eq. (\ref{sh2}),
the {\it absolute} entropy of a continuous variable is not well defined
since it diverges in the limit $\Delta x\rightarrow 0$.
Nevertheless, the {\it entropy
difference := information} is well-defined and {\it does not} 
depend on the precision $\Delta x$.

\subsection{$\tau$-Information}

The generalization of  information theory
onto the case of stochastic processes is not trivial.
In our case, the proper definition of entropy of the switch-point process
{$\tau (\tilde t) $}, considered on the time interval $[0,T]$, is -- by
analogy with Eq. (\ref{sh2}) --
\begin{eqnarray}\label{tau-S}
S_{\tau}[T|V_s]  && \equiv   -\kappa\sum_{\alpha= o,c}\Big \{Q_0^{\alpha}(T)
\ln Q_0^{\alpha}(T) \nonumber \\  && + 
\sum_{s=1}^{\infty}
\int_{0}^{T}d\tau_s  \int_{0}^{\tau_s}d\tau_{s-1}...
\int^{\tau_2}_{0}d\tau_1
Q^{\alpha}_s(T,\tau_s,..,\tau_1)\nonumber \\ && \times 
\ln [Q_s^{\alpha}(T,\tau_s,..,\tau_1)
(\Delta\tau)^s]
\Big\}\;,
\end{eqnarray}
where $\Delta\tau$ denotes the  precision of time measurement, and the
symbol $V_s$ indicates
that the entropy is defined in  presence of the signal $V_s(t)$.
The presence of the  time resolution
$\Delta\tau$ in (\ref{tau-S}) gives the name ``$\tau$-entropy'' to this
quantity \cite{gaspard}. It is very important that in the contrast to the case
of a continuous variable, the contribution of the finite time resolution
$\Delta\tau$
to the $\tau$-entropy cannot be recasted in a form like $-\kappa\ln\Delta x$, 
cf. Eq. (\ref{sh2}). We note that its contribution {\it depends on the
statistics of
the random process}, being different in the presence and in the absence of
signal. This is  why not only the {\it absolute} entropy,
but also the {\it difference} of entropies become not well defined
for continuous time point random processes. As a result, the
definition of information in this manner becomes rather ambiguous.

For a sufficiently large time interval
$T$ the averaged information transferred per unit time from the
input voltage signal $V_s(t)$ to the output current signal $I(t)$
can be defined as follows\cite{bialek,strong}
\begin{eqnarray}\label{t-inf}
{\cal I}_{\tau}=\frac{S_{\tau}(T|V_s=0)-S_{\tau}(T|V_s)}{T}.
\end{eqnarray}
This information measure can be termed $\tau$-information per unit time
to underline its dependence on the time resolution $\Delta\tau$.
Upon taking the derivative of $S_{\tau}[t|V_s]$ in
Eq. (\ref{tau-S}) with respect to time $t$ we obtain after some
 involved algebra (cf. Appendix B) the result
\begin{eqnarray}\label{result1}
\frac{dS_{\tau}[t|V_s]}{dt}=-\kappa\sum_{\alpha=o,c}\bar k_{\alpha}(t)\ln
\Big(\bar k_{\alpha}(t)\Delta\tau/e \Big)P_{\bar\alpha}(t)\;,
\end{eqnarray}
where $\bar \alpha=o$, if $\alpha=c$ and {\it vice versa}.
Together with Eq. (\ref{balance}) and the definition (\ref{t-inf})
the prominent result in
Eq. (\ref{result1}) allows one to express the
$\tau$-information for arbitrary signal $V_s(t)$
through straightforward quadratures.

The $\tau$-information concept has been used in fact to analyze the information
transfer in neuronal systems in Ref. \cite{bialek,strong}. However, the
strong dependence of $\tau$-information on the time
precision $\Delta\tau$ \cite{strong} presents surely an undesirable
{\it subjective} feature.
In search
for {\it objective} information measures we
consider the
information transfer in terms of the  mutual information measure.

\subsection{Mutual information}

To introduce the reader to the mutual information concept,
we follow the reasoning of Shannon
\cite{shannon}:  the signals $V_s(t)$ are
drawn from some statistical distribution
characterized by the probability density functional $P[V_s(t)]$.
Noting that the probability densities $Q_{s}^{\alpha}
(t,\tau_s,...,\tau_1)$
in Eqs. (\ref{Q0}), (\ref{qeven}) and (\ref{qodd}) are in fact
{\it conditional} with respect to the given realization of $V_s(t)$,
one can define the joint
probability densities,
$Q_{joint}^{\alpha,s}(t,\tau_s,...,\tau_1;V_s(t))=
Q_{s}^{\alpha}(t,\tau_s,...,\tau_1)
P[V_s(t)]$ for the corresponding stochastic processes $V_s(t)$ and $I(t)$.
Moreover, one can define the
averaged probability densities
$\langle Q_{s}^{\alpha}(t,\tau_s,...,\tau_1)\rangle_{signal}$ for the
process $I(t)$ {\it in the presence of the process} $V_s(t)$,
where the path integral $\langle...\rangle_{signal}\equiv \int {\cal D} V_s(t)
... P[V_s(t)]$
denotes stochastic averaging over the signal realizations.
The mutual information between the stochastic process
$V_s(t)$ and $I(t)$ can then be defined as the entropy
difference
\begin{eqnarray}\label{mi}
{\cal M}_T(V_s,I)=S_{av}(T)-\langle S_{\tau}(T|V_s) \rangle_{signal},
\end{eqnarray}
where $S_{av}(T)$ is the $\tau$-entropy of the averaged process
defined similarly to (\ref{tau-S}), but with
the {\it averaged} probability densities
$\langle Q_{s}^{\alpha}(t,\tau_1,...,\tau_s)\rangle_{signal}$. Note
that making use of the Bayes rules one can transform the definition
(\ref{mi}) into
a form which makes transparent the fact that the mutual information
${\cal M}_T(V_s,I)$ is a symmetric functional
of the processes $V_s(t)$ and $I(t)$ and provides a {\it nonlinear}
cross-correlation measure between them \cite{shannon}. We, however, will
take advantage
of an equivalent form; it is obtained from Eq. (\ref{mi}) by
using Eq. (\ref{tau-S}), yielding
\begin{eqnarray}\label{mi2}
&& {\cal M}_T (V_s,I) = \kappa\Big\langle \sum_{\alpha=o,c}\Big\{Q_0^{\alpha}(T)
\ln\frac{Q_0^{\alpha}(T)}{\langle Q_0^{\alpha}(T)\rangle_{signal}} 
\nonumber \\ & + &
\sum_{s=1}^{\infty}
\int_{0}^{T}d\tau_s\int_{0}^{\tau_s}d\tau_{s-1}...
\int^{\tau_2}_{0}d\tau_1
Q^{\alpha}_s(T,\tau_s,..,\tau_1)\nonumber \\ && \times
\ln\frac{Q_s^{\alpha}(T,\tau_s,..,\tau_1)}
{\langle Q_s^{\alpha}(T,\tau_s,..,\tau_1)\rangle_{signal}}
\Big \} \Big\rangle_{signal}, 
\end{eqnarray}
As is clearly deduced from Eq. (\ref{mi}),
 Shannon's mutual
information {\it does not}
depend  -- due to its skillful definition in  Eq. (\ref{mi}) --  on the
time resolution $\Delta\tau$. This underpins its
advantage over the information measure in Eq. (\ref{t-inf}). Moreover,
the functional form (\ref{mi2}) inherits important connections between
the mutual information
and another prominent information measure -- the (relative) Kullback entropy or
termed also the {\it information gain}.

\subsection{Rate of Information gain}

The information gain  \cite{beck}
is given in terms of the  relative entropy of
the given statistical
distribution with respect to some reference distribution. In our case,
the reference distribution corresponds to the stationary
ion current fluctuations in the absence of the  voltage signal $V_s(t)$.
For a given signal $V_s(t)$ the information gain reads
\begin{eqnarray}\label{kul}
&&{\cal K}_T[I|V_s]  \equiv  \kappa\sum_{\alpha=o,c}\Big\{Q_0^{\alpha}(T)
\ln\frac{Q_0^{\alpha}(T)}{Q_0^{(0)\alpha}(T)} \nonumber \\ & + &
\sum_{s=1}^{\infty}
\int_{0}^{T}d\tau_s\int_{0}^{\tau_s}d\tau_{s-1}...
\int^{\tau_2}_{0}d\tau_1
Q^{\alpha}_s(T,\tau_s,..,\tau_1)\nonumber \\ && \times
\ln\frac{Q_s^{\alpha}(T,\tau_s,..,\tau_1)}
{Q_s^{(0)\alpha}(T,\tau_s,..,\tau_1)}
\Big\}\;,
\end{eqnarray}
where the index $``(0)"$ in $Q_s^{(0)\alpha}$ refers to the case
when no voltage signal is applied. The relative entropy can be regarded as the
signal-induced
deviation of entropy of the random point process $\tau(\tilde t)$ from its
stationary value
obtained in the absence of signal.
Although the {\it absolute} entropy of such a switch-time point  process
$\tau(\tilde t)$ depends
strongly on the time resolution $\Delta\tau$ and thus is not well-defined,
the deviation of
entropy from the steady-state value can be defined {\it independently}
of $\Delta\tau$ via Eq. (\ref{kul}).
For stochastic processes this relative entropy
plays the role similar to the entropy difference; thus characterizing
an information measure. This justifies its given name --
the information gain. In contrast to  mutual information
this measure can be defined
 for {\it deterministic} signals as well. Consequently, the information
gain can be used as an information
measure both for conventional and for aperiodic SR. Moreover, this
measure is also well-defined for {\it nonstationary} signals and therefore
can be used
to quantify  {\it nonstationary} SR as well.

In contrast to the information gain the mutual information is more
difficult  to handle
analytically. This  is rooted in the fact
 that the
{\it averaged}
point process {$\tau(\tilde t)$} is non-Markovian process  with
corresponding joint probabilities not factorizing into products of
conditional probabilities.

The following important inequality can
be deduced
\begin{eqnarray}
\label{ineq}
{\cal M}_T(V_s,I)&& =\langle  {\cal K}_T[I|V_s]
\Big\rangle_{signal}-{\cal K}_T[\langle I\rangle_{signal}]\nonumber \\ && 
\leq \Big\langle  {\cal K}_T[I|V_s]
\Big\rangle_{signal}\;.
\end{eqnarray}
In Eq. (\ref{ineq}), ${\cal K}_T[\langle I\rangle_{signal}]\geq 0$ is
the relative entropy of the {\it averaged} process $g(t)$ defined 
similarly to Eq. (\ref{kul}), but with 
the averaged multi-time probability densities 
$\langle Q_s^{\alpha}(T,\tau_s,..,\tau_1)\rangle_{signal}$.
The averaged information gain provides thus an upper bound for
the mutual information.  Moreover, applying a weak Gaussian signal
in the limit $\tau_s/\tau_I \rightarrow 0$  one can show that the
difference between
the mutual information and the averaged information gain in (\ref{ineq})
is of order $O(A^4)$, where $A$ denotes the r.m.s. amplitude of signals
$A=\langle V^2_s(t)\rangle_{signal}^{1/2}$. On the other hand,  it is shown below
that
the averaged information gain per unit time is  of the order $O(A^2)$
and does not depend, within the given lowest order approximation,
on other statistical parameters of signal. Thus, the upper bound
for mutual information in Eq. (\ref{ineq}) can indeed be achieved 
with an accuracy of $O(A^2)$. 
This fact opens a way to calculate the informational capacity
for weak signals \cite{remark2}.

The information
gain can be evaluated from Eq. (\ref{kul}) without further
problems. By differentiating
${\cal K}_T[I|V_s]$  with respect to $T$
 we find following to the reasoning detailed in 
 Appendix B the remarkable simple, {\it main} result for the
{\it rate of
information gain}, i.e.
\begin{eqnarray}\label{eq1}
\frac{d{\cal K}_t[I|V_s]}{dt}=&&\kappa\sum_{\alpha=o,c}[\bar k_{\alpha}(t)\ln
\Big(\frac{\bar k_{\alpha}(t)}{\bar k_{\alpha}(V_0)} \Big)\nonumber \\ && 
-\bar k_{\alpha}(t)
+\bar k_{\alpha}(V_0)]P_{\bar\alpha}(t)\;,
\end{eqnarray}
wherein $\bar k_{\alpha}(V_0)$ denote the stationary transition rates
in the absence of signal.
Together with Eq. (\ref{balance}) this equation {\it completely} determines
the information gain within the considered two-state model for any
applied signal $V_s(t)$. For the case of a periodic signal $V_s(t)$
(conventional SR),
or a stochastic stationary signal (aperiodic SR), one should
average additionally Eq. (\ref{eq1}) over the signal fluctuations and to
take the limit
$t\rightarrow \infty$. In doing so, Eq. (\ref{eq1}) yields the stationary
rate of  information gain. For  weak stochastic signals this
quantity also defines the  informational capacity \cite{remark2}

\begin{equation}
{\cal C}\approx
\lim_{T\rightarrow\infty}\Big\langle  {\cal K}_T[I|V_s]
\Big\rangle_{signal}/T \;.
\end{equation}

 If the signal is deterministic and has a finite duration, one obtains the {\it
total} information gain ${\cal K}$ by integrating Eq. (\ref{eq1}) in the range
from $0$ to $\infty$.

\section{Stochastic resonance  in single K$^+$ ion channels}

In the following we apply our developed information theory concepts to
investigate SR
in a K$^+$ ion channel. We restrict our treatment  to the case of  weak
signals with a time duration which strongly exceeds  the autocorrelation
time of current fluctuations $\tau_{I}$.
Then, Eqs. (\ref{eq1}) and (\ref{balance})
yield after some elementary calculations in the lowest order of $V_s(t)$,
\begin{eqnarray}
\label{res1}
\frac{d{\cal K}_t[I(t)|V_s(t)]}{dt}=R(V_0,\sigma) V_s^2(t)
\end{eqnarray}
where the form factor
\begin{eqnarray}
\label{res2}
R(V_0,\sigma)=\frac{1}{8}\,\kappa \,
\frac{\bar k_o(V_0)\bar k_c(V_0)}{\bar k_o(V_0)+\bar k_c(V_0)}
\Big [\beta_o^2(V_0)+\beta_c^2(V_0) \Big ],
\end{eqnarray}
depends -- via the rates $\bar
k_{\alpha}(V_0)$ -- on the static voltage $V_0$ and on the  r.m.s.
noise amplitude $\sigma$.
In Eq.(\ref{res2}), $\beta_{\alpha}(V_0)=2\frac{d}{dV_0}\ln[ \bar
k_{\alpha}(V_0)],\;
\alpha=o,c$,
and the noise averaged rates $\bar k_{o(c)}(V_0)$ are given in the Appendix A
for a K$^+$ channel in Eqs. (\ref{r1}) and (\ref{r2}).

In the case of stationary stochastic signals or for a periodic driving,
 Eq. (\ref{res1}) provides after stochastic averaging, or averaging
over the driving period of applied voltage $V_s(t)$, respectively, the
stationary rate of  information gain. For signals
of finite duration the total information gain is directly proportional to
the total intensity of signal $\xi=\int_0^{\infty}V_s^2(t)dt$,
\begin{eqnarray}
\label{res3}
{\cal K}=R(V_0,\sigma) \xi.
\end{eqnarray}
As a result we find that weak signals of the
 the same intensity $\xi$ produce equal information gains. The
occurrence of three different kinds of SR behavior, i.e., periodic,
aperiodic, and nonstationary SR clearly depends on the
behavior of
the form function
$R(V_0,\sigma)$ {\it vs.} the r.m.s. noise amplitude $\sigma$. We recall
that the
static voltage (membrane potential) $V_0$ controls whether
the ion channel is  on average open or closed, cf. Fig.\ref{Fig1}. In Fig.
\ref{Fig2}, we depict the behavior of the function $R(V_0,\sigma)$
{\it vs.} the  r.m.s. noise amplitude for different values of the applied
static voltage.
If the K$^+$  ion channel is closed on average we observe that
the information gain
becomes strongly be amplified by  noise, and even  can pass through a maximum,
i.e. SR occurs , cf. Fig.2a. In contrast, when the stationary
probability for an open channel $P_o=k_o/(k_o+k_c)$
becomes appreciably large, the addition of an additional dose of noise
can only deteriorate the detection of signal.
As  a result, the information gain decreases
monotonically with  increasing
noise amplitude, cf. Fig. 2b. This {\it no}\/-SR behavior occurs already at  a
static bias of $V_0\approx -49$ mV yielding
$P_o\approx 0.08$. Note also, if the channel is predominantly open,
the information gain becomes practically insensitive to the external
noise, 
cf. the bottom curve in Fig. 2b. Although the information gain
can slightly be increased versus the noisy intensity in this case
($V_0=-20\;{\rm mV}$), 
this effect is hardly of importance because the overall information
gain diminishes drastically with increasing the static
voltage $V_0$ (see Fig. 2b). The occurrence of SR in 
the considered single ion
channel thus
requires that the channel is predominantly resting in its closed state.

\section{Conclusions}

Let us now summarize  the main results of this work. We have studied an
illustrative
two-state model for  a single ion channel  gating dynamics from an
information theoretic
point of view. The channel serves as
 an information channel transducing information
from the applied time-dependent voltage
signal to the ion current fluctuations. Three
different
information theory measures have   been developed to characterize
stochastic resonance. From
our viewpoint it is advantageous to use
an information measure which is independent of
 time resolution $\Delta\tau$. We argued
that the rate of information gain  constitutes a unified
characteristic measure for periodic (conventional), aperiodic and
nonstationary stochastic resonance.
For conventional (periodic) SR and aperiodic SR this measure
yields the averaged information gain per unit time. Moreover,
for weak stochastic signals it gives also
the informational capacity, i.e., the maximal mutual information
which can be transferred per unit time for random signals with a fixed
r.m.s. amplitude. The concept of information gain can also
be applied to the case of {\it nonstationary} deterministic
signals with finite duration, i.e., nonstationary SR, cf. Eqs. (\ref{res2}),
(\ref{res3}).

Our main result is the closed formula for the rate of  information gain in
(\ref {eq1}):
 it can be evaluated
 in a straightforward manner by using the corresponding
probabilities of the two-state gating dynamics in (\ref{balance}). The
information gain itself follows
upon a time integration. In presence of weak driving
we derived handy analytical results given in
Eqs. (\ref{res1}), (\ref{res2}), and (\ref{res3}). For   voltage input signals
referring to a stationary process the averaged rate of the information
gain is determined by the r.m.s. amplitude of the  signal input
and by the form factor $R(V_0,\sigma)$. In the case of a nonstationary
signal of finite duration, the total information gain is the product of
this very form function
 and the integrated signal
intensity~ $\xi$.

The experimental procedure of determining the rate of
information gain can be formulated along the lines used for the $\tau$-entropy
in Ref. \cite{strong}.
First, one finds the corresponding probability histograms in the
presence and in the absence of signal and then evaluates the information
gain for the related binary stochastic chains. Naturally, this so obtained
information gain  will still depend on the
time resolution $\Delta\tau$.
However, in contrast to the $\tau$-information, this experimentally
determined information gain
should exhibit a much weaker dependence on the time resolution $\Delta\tau$.
By using increasingly smaller time grids $\Delta\tau$, the experimentally
obtained rate of information gain will  approach a definite  value.

Our theoretical results have been applied to investigate the phenomenon of
stochastic resonance
 in a potassium-selective {\it Shaker IR} ion channel
\cite{salman}, as depicted with in Fig. 2a,b. Interestingly enough, we find
that periodic, aperiodic or nonstationary SR for this sort of ion channel,
as quantified by the rate of information gain,
is exhibited only for a situation in which the channel resides on average
in the closed state. This type of behavior is rooted in the asymmetry
of two rates $k_o$ and $k_c$; with $k_o$ depicting a characteristic
steep, threshold-like behavior, cf. Fig. 1.

Our SR-feature is   similar to  the study of parallel SR in an array of
alamethicin channels
 \cite{bez}, although the two situations are, however, not directly comparable.
We note that
the amount of transmitted information crucially depends  on the
membrane potential $V_0$. For the studied model, the information
transfer is optimized at zero noise level
near  $V_0\approx -46$ mV when the opening probability becomes appreciable
(note the upper curve in Fig. 2b). However, under such optimal conditions
the addition of external noise has the effect of only further deteriorating
the rate of information
transfer (Fig. 2b). Upon   further increasing the static bias $V_0$ the ion
channel
 probability to stay open increases. The rate of  information transfer then
diminishes
 and becomes practically insensitive to the input noise level.

These results hopefully will motivate researchers to measure the predicted
SR behavior
 in single potassium ion channels. Ever since the discovery of the SR
phenomenon,
the quest to use noise to optimize and control
the transduction and relay of biological information has been one of the
Holy Grails
  of SR research. Given this challenge, such and related experiments  are
much needed
in order to settle the issue in question.

\acknowledgments
The authors gratefully acknowledge the support of this work
by the German-Israel-Foundation G.I.F. G-411-018.05/95, as well as
by the Deutsche Forschungsgemeinschaft (SFB 486 and HA1517/13-2).

\appendix
\section{Model for averaged transition rates in a potassium channel }

The opening and closing rates for the effective two-state
model can be found from the voltage-dependent average dwell times.
The latter can be   determined from the experimental recordings.
The experimental dependence of the effective transition rates
on voltage $V_0$ for the potassium-selective channel {\it Shaker IR}
embedded in
the membrane
of a {\it Xenopus} oocyte at  {\it fixed} temperature $T=18\;^oC$
have been fitted \cite{marom,salman} by a Hodgkin-Huxley type of data
parameterization \cite{hodgkin}.
This corresponding fitting procedure 
 yields
\begin{eqnarray}\label{rates}
k_c(V)& = &a_1 e^{-b_1V},\;\;a_1=0.015, \;\;b_1=0.038, 
 \\
k_o(V)& = & \frac{a_2(V+V_2)}{1-e^{-b_2(V+V_2)}},\;
\;a_2=0.03, \;b_2=0.8,\;V_2=46,\nonumber
\end{eqnarray}
which are  depicted in Fig. \ref{Fig1}. Note that we replace the
original fit of the closing rate $k_c$ in \cite{marom,salman} by
a new expression in Eq. (A1). Unlike to \cite{marom}, our fit
of experimental data in \cite{marom} is valid now also for positive
voltages $V$. One should emphasize,
that the two rates in Eq. (A1) are strongly
asymmetric with respect to their dependence on voltage. In particular,
the opening rate $k_o(V)$ depicts a steep, threshold-like behavior, see Fig. 1. 
In this work we explicitly use
these experimental
findings in our calculations. The
rates in Eq. (\ref{rates}) are measured in ${\rm msec}^{-1}$ and the
voltage in mV. According to our model study, the input voltage reads
$V=V_0+V_n(t)$
when no additional signal is applied. These Eqs.
(\ref{rates}) must be averaged over the realizations of $V_n(t)$ to obtain
the noise
averaged rates $\bar k_o(V_0)$ and $\bar k_c(V_0)$.
For a Gaussian  voltage noise $V_n(t)$ this averaging of the exponential in the
first equation in (\ref{rates}) is governed by the second cumulant, yielding
\begin{eqnarray}\label{r1}
\bar k_c(V_0) = a_1 e^{b_1^2\sigma^2/2-b_1V_0},
\end{eqnarray}
where $\sigma=\langle V_n^2(t)\rangle^{1/2}$.
The averaged opening rate
\begin{eqnarray}\label{r2}
\bar k_o(V_0)=\frac{1}{\sqrt{2\pi}\sigma}\int_{-\infty}^{\infty}
\frac{a_2(V_0+V_2+y)}{1-e^{-b_2(V_0+V_2+y)}}e^{-\frac{y^2}{2\sigma^2}}dy
\end{eqnarray}
unfortunately cannot be analytically simplified further. However,
this rate along with its derivative $d \bar k_o(V_0)/dV_0$ can
readily be evaluated numerically from Eq. (\ref{r2}).

\section{Calculation of Entropy and Information Gain}

The purpose of this Appendix to provide the readers with some details of 
calculation of the entropic measures for the continuous time random 
point two-state process considered in this paper. First,
we note two useful properties of the multi-time probability densities
which can be established from Eqs. (\ref{qeven}), (\ref{qodd}). 
Namely,
\begin{eqnarray}\label{a1}
\frac{d}{dt}Q^{\alpha}_s(t,\tau_s,..,\tau_1)=-\bar k_{\bar\alpha}(t)
Q^{\alpha}_s(t,\tau_s,..,\tau_1), \;s\geq 0,
\end{eqnarray}
and
\begin{eqnarray}\label{a2}
Q^{\alpha}_s(t,t,\tau_{s-1},..,\tau_1)=\bar k_{\alpha}(t)
Q^{\bar\alpha}_{s-1}(t,\tau_{s-1},..,\tau_1), \;s\geq 1.
\end{eqnarray}
The index $\alpha$ in Eqs. (\ref{a1}), (\ref{a2}) takes the values 
$\alpha=o,c$; and the index $\bar\alpha$ takes the value $\bar \alpha=o$, 
if $\alpha=c$, and {\it vice versa}. Using Eqs. (\ref{a1}), (\ref{a2}) 
one can check that $P_{o(c)}$ given in Eq. (\ref{solution}) do 
satisfy Eq. (\ref{balance}). 

Furthermore, let us consider the
$\tau$-entropy in Eq. (\ref{tau-S}) as a sum of two contributions, 
$S_{\tau}[t|V_s]=\kappa\sum_{\alpha=o,c}S_{\alpha}(t)$,
with $S_{\alpha}(t)$ defined from the corresponding 
partitioning in Eq. (\ref{tau-S}). 
Then, using repeatedly the relationships (\ref{a1})
and (\ref{a2}) we obtain after some straightforward, but lengthy calculations
\begin{eqnarray}\label{a3}
\frac{d}{dt}S_o(t) = &-&\bar k_c(t)S_o(t)+\bar k_o(t)S_c(t) \nonumber \\
& + &\bar k_c(t)P_o(t)-\bar k_o(t)\ln(\bar k_o(t)\Delta \tau)P_c(t)
\end{eqnarray}
and
\begin{eqnarray}\label{a4}
\frac{d}{dt}S_c(t) = &-&\bar k_o(t)S_c(t)+\bar k_c(t)S_o(t)\nonumber \\
& + &\bar k_o(t)P_c(t)-\bar k_c(t)\ln(\bar k_c(t)\Delta \tau)P_o(t).
\end{eqnarray} 
The addition of Eq. (\ref{a3}) and Eq. (\ref{a4}) then yields
Eq. (\ref{result1}). Likewise, splitting the 
information gain ${\cal K}_t[I|V_s]$ in 
Eq. (\ref{kul}) into the sum of two contributions, ${\cal K}_t[I|V_s]=
\kappa\sum_{\alpha=o,c} K_{\alpha}(t)$, and invoking the properties  
(\ref{a1}) and (\ref{a2}) we obtain after some algebra 
\begin{eqnarray}\label{a5}
\frac{d}{dt}K_o(t)= && -\bar k_c(t)K_o(t)+\bar k_o(t)K_c(t)\nonumber \\
&& - [\bar k_c(t)-\bar k_c(V_0)]P_o(t)\nonumber \\ && 
+\bar k_o(t)\ln\Big (\frac{\bar k_o(t)}
{\bar k_o(V_0)}\Big)P_c(t)
\end{eqnarray}
and 
\begin{eqnarray}\label{a6}
\frac{d}{dt}K_c(t)= && - \bar k_o(t)K_c(t)+\bar k_c(t)K_o(t)\nonumber \\
&& - [\bar k_o(t)-\bar k_o(V_0)]P_c(t)\nonumber \\ && 
+\bar k_c(t)\ln\Big (\frac{\bar k_c(t)}
{\bar k_c(V_0)}\Big)P_o(t)\;.
\end{eqnarray}
Adding Eq. (\ref{a4}) and Eq. (\ref{a5}) results 
after multiplying with $\kappa$ in our main result in Eq. (\ref{eq1}).

\newpage
\begin{figure}[h]
\begin{center}
\leavevmode
\epsfxsize=0.45\textwidth
\epsfbox{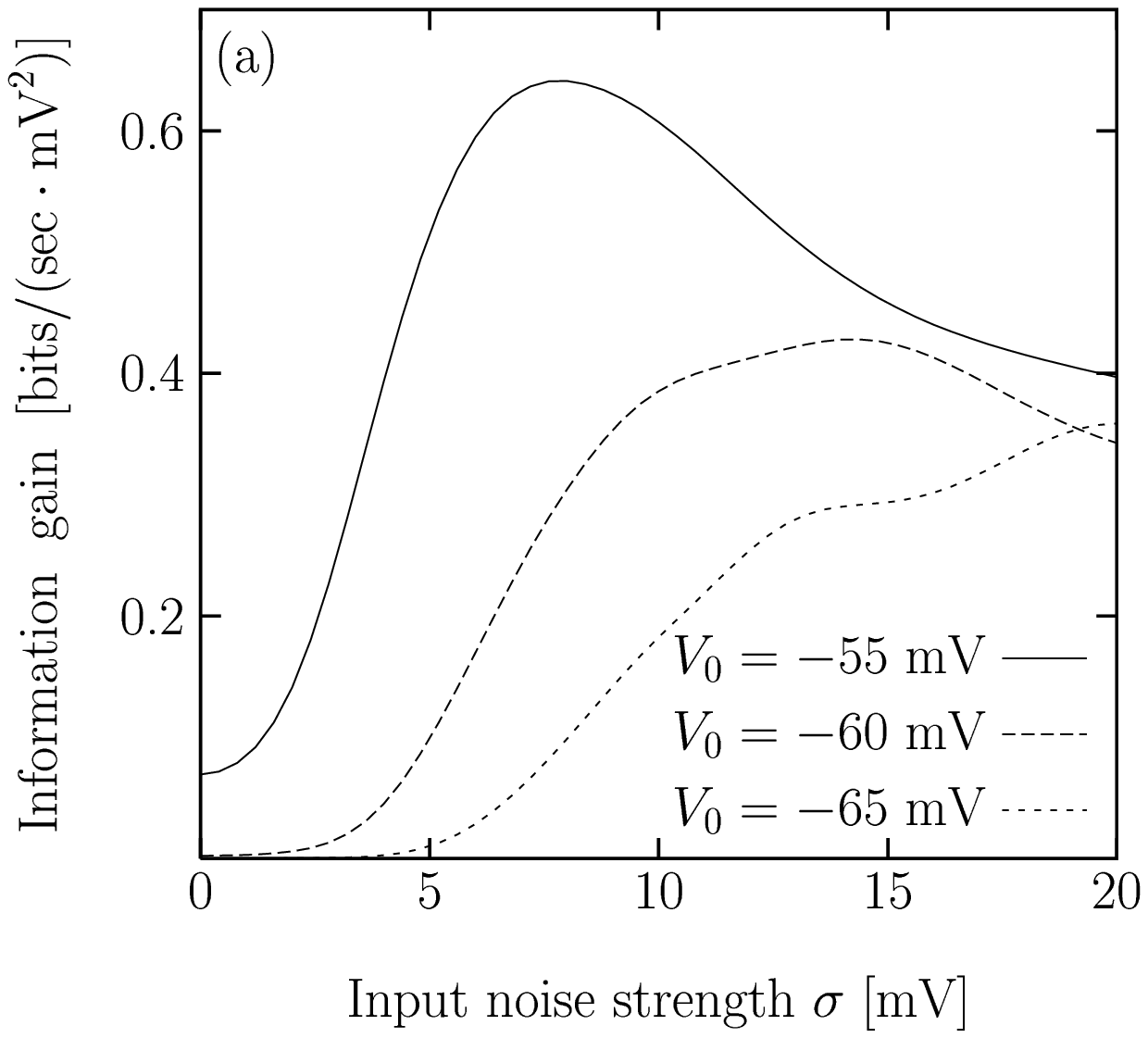}
\end{center}
\begin{center}
\leavevmode
\epsfxsize=0.45\textwidth
\epsfbox{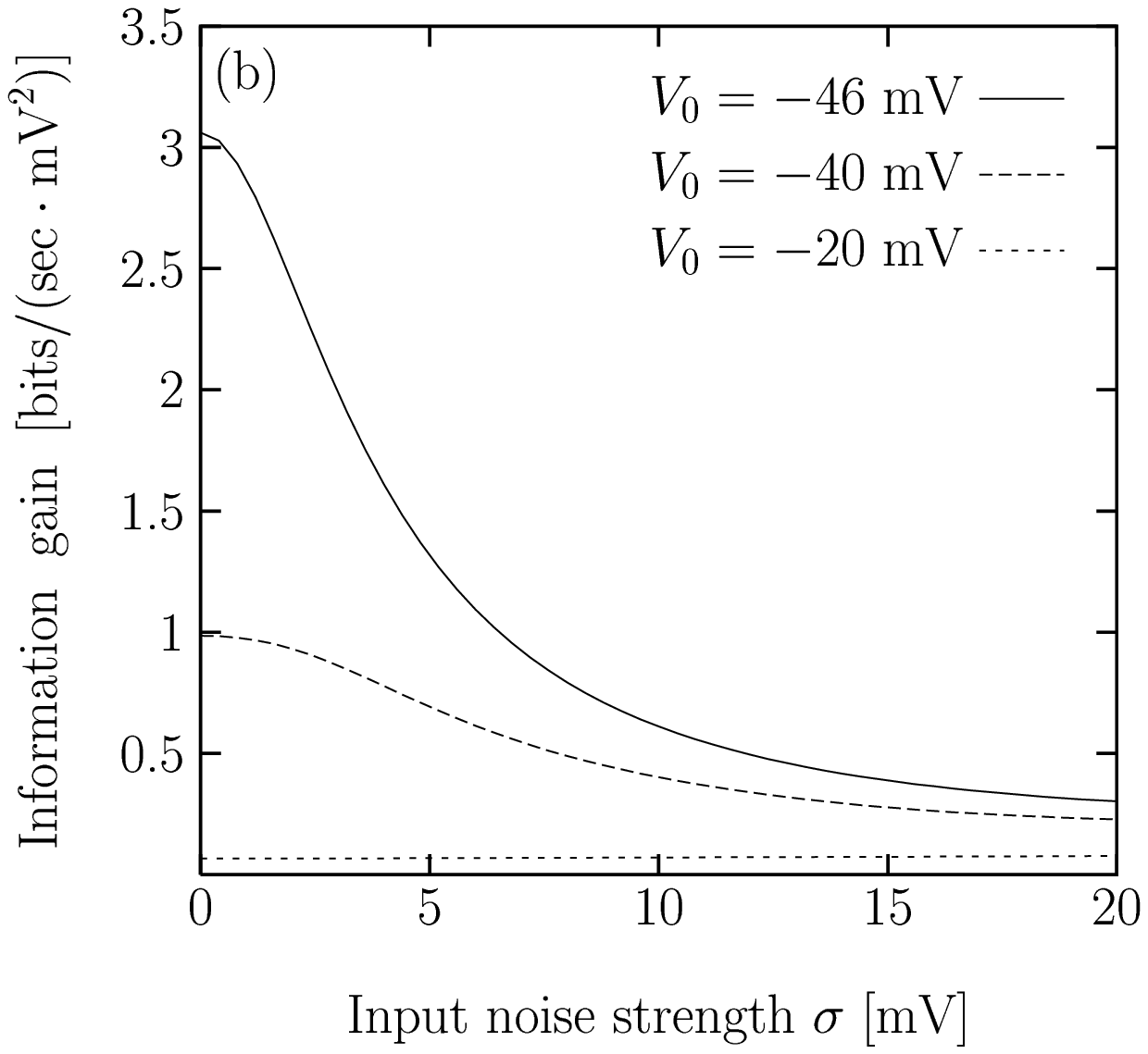}
\end{center}
\end{figure}
\end{multicols}
\begin{figure}
\caption{Information gain versus r.m.s. amplitude of external noise
$\sigma$ at various static bias voltages $V_0$. The form function
$R(V_0,\sigma)$
in Eqs. (19)-(21) is plotted vs. the r.m.s. noise intensity $\sigma$.}
\label{Fig2}
\end{figure}

\begin{references}
\bibitem{wies} K. Wiesenfeld and F. Jaramillo, Chaos {\bf 8}, 539 (1998).
\bibitem{rev1} L. Gammaitoni, P. H\"anggi, P. Jung, and F. Marchesoni,
Rev. Mod. Phys. {\bf 70}, 223 (1998).
\bibitem{longtin} A. Longtin, A. Bulsara, and F. Moss, Phys. Rev. Lett. {\bf 67},
656 (1991).
\bibitem{sim}E. Simonotto, {\it et al.}, Phys. Rev. Lett. {\bf 78}, 1186
(1997).
\bibitem{col1}J.J. Collins, T.T. Imhoff, and P. Grigg, Nature (London)
{\bf 383}, 770 (1996).
\bibitem{levin}J. E. Levin, J. P. Miller, Nature (London) {\bf 380}, 165
(1996).
\bibitem{gluck} B.J. Gluckman, P. So, T.I. Netoff, M.L. Spano, and
S.J. Schiff, Chaos {\bf 8}, 588 (1998).
\bibitem{doug} J.K. Douglass, L.Wilkens, E. Pantazelou, and F. Moss,
Nature (London) {\bf 365}, 337 (1993).
\bibitem{hille}B. Hille, {\it Ionic Channels of Excitable Membranes}, 2nd ed.
(Sinauer Associates, Sunderland, MA, 1992).
\bibitem{bez}S.M. Bezrukov and I. Vodyanoy, Nature (London) {\bf 378}, 362
(1995).
\bibitem{bez2}S.M. Bezrukov and I. Vodyanoy, Nature (London) {\bf 385}, 319
(1997); S.M. Bezrukov, Phys. Lett. A {\bf 248}, 29 (1998).
\bibitem{collins} J. J. Collins, C. C. Chow, and T. T. Imhoff, Nature
(London) {\bf 376}, 236 (1995);
J. J. Collins, C. C. Chow, and T. T. Imhoff, Phys. Rev. E
{\bf 52} R3321 (1995); J. J. Collins, C. C. Chow, A. C. Capela
and T. T. Imhoff, Phys. Rev. E {\bf 54} 5575 (1996); C. Heneghan,
C. C. Chow, J. J. Collins, T. T. Imhoff, S. B. Lowen, and M. C. Teich,
 Phys. Rev. E {\bf 54}, R2228 (1996).
 \bibitem{stemmler} M. Stemmler, Network {\bf 7},687 (1996).
\bibitem{buls} A. D. Bulsara and A. Zador, Phys. Rev. E {\bf 54},
R2185 (1996).
\bibitem{shannon} K. Shannon, Bell System Technical Journal {\bf 27},
pp. 379-423, 623-656 (1948).
\bibitem{kramers}H. A. Kramers, Physica {\bf 7}, 284 (1940).
\bibitem{hanggi1}P. H\"anggi, P. Talkner, and M. Borkovec, Rev. Mod. Phys.
{\bf 62}, 251 (1990).
\bibitem{hanggi2} P. Reimann and P. H\"anggi, in: Springer-Series {\it
Lecture Notes in Physics}, ed. by L. Schimansky-Geier and T. P\"oschel
(Springer, Berlin, 1997), Vol. 484, pp. 127-139.
\bibitem{bezanilla} D. Sigg, H. Qian, and F. Bezanilla, Biophys. J. {\bf 76},
782 (1999).
\bibitem{sak}B. Sakmann and E. Neher (eds.), {\it Single-Channel Recording},
2nd ed. (Plenum, New York, 1995).
\bibitem{salman} H. Salman, Y. Soen, and E. Braun, Phys. Rev. Lett. {\bf
21},4458.
(1996); H. Salman and E. Braun, Phys. Rev. E {\bf 56}, 852 (1997).
\bibitem{marom}S. Marom, H. Salman, V. Lyakhov, and E. Braun,
J. Membrane Biol. {\bf 154}, 267 (1996).
\bibitem{mc} B. McNamara and K. Wiesenfeld, Phys. Rev. A {\bf 39}, 4854 (1989).
\bibitem{neiman}A. Neiman, L. Schimansky-Geier, and F. Moss,
Phys. Rev. E {\bf 56}, R9 (1997).
\bibitem{remark1}The problem of extracting these conductance fluctuations
from the current recordings in the presence of a time-dependent
(e.g., periodic) driving is explained in:
D. Petracchi {\it et al.}, J. Stat. Phys. {\bf 70}, 393 (1993).
\bibitem{kamp}N.G. Van Kampen,{\it Stochastic Processes in Physics and
Chemistry, 2-d, enlarged and extended edition} (North-Holland, Amsterdam,
1992).
\bibitem{strat}R. L. Stratonovich, {\it Topics in the Theory
of Random Noise}, vol. I (Gordon and Breach, New York, 1963).
\bibitem{remark} It is remarkable that the permutation invariance of $S({\cal
A})=S(p_1,p_2,...,p_n)$ with respect to
 the set of probabilities ${p_i}$
and the property of additivity, i.e. $S({\cal A} \times {\cal B}) = S({\cal
A}) + S({\cal B})$ --
in case that the probabilities factorize in the composed state space,
 or subadditivity, in case of statistical dependence  of two
composed (i.e. Cartesian product) measure spaces -- characterize the
 Shannon entropy $S({\cal A}) = S(p_1,...,p_n)$ almost uniquely: any
functional satisfying
these requirements is a
linear combination of the Shannon entropy and the Hartley entropy ($ = \ln
{n^*}$,
with $ n^* $ being the number of ${p_i{\rm '}s}$ that are different from zero).
The additional requirements
of (i) $S(p,1-p)$ being a continuous function of $p$, $0 \leq p \leq 1$,
and (ii)
 $S(tp_1,[1-t]p_1,p_2,...,p_n) = S(p_1,....,p_n)  + p_1 S(t,1-t)$, with $0
\leq t \leq 1$
(A. Feinstein, {\it Foundations of Information Theory}\/, (Mc Graw Hill, New
york, 1958))
determine then the Shannon entropy uniquely.
\bibitem{gaspard}P. Gaspard and X.-J. Wang, Phys. Rep. {\bf 235}, 292 (1993).
\bibitem{bialek}F. Rieke, D. Warland, R. de Ruyter van Steveninck,
and W. Bialek, {\it Spikes: Exploring the Neural Code}
(MIT Press, Cambridge, MA, 1997).
\bibitem{strong} S. P. Strong, R. Koberle, R. R. de Ruyter van Steveninck,
and W. Bialek, Phys. Rev. Lett. {\bf 80}, 197 (1997).
\bibitem{beck} The many facets of entropy are beautifully outlined in: A.
Wherl,
Rep. Math. Phys. {\bf 30},119 (1991); see also: C. Beck and F. Schl\"ogl,
{\it Thermodynamics of Chaotic
Systems:
an Introduction}, (Cambridge University Press, Cambridge, 1993).
\bibitem{remark2} The informational capacity of an informational
channel is defined as the maximal rate of mutual information obtained 
for all possible statistical distributions of input signals with {\it fixed}
r.m.s. amplitude \cite{shannon}.
 \bibitem{hodgkin}A. L. Hodgkin and A. F. Huxley, J. Physiol. {\bf 117}, 500
(1952).
\end{references}
\end{document}